\documentclass{webofc}
\usepackage[varg]{txfonts}   
\usepackage{booktabs}
\usepackage{amsmath,amssymb}
\usepackage{adjustbox}
\usepackage{multirow}
\newcommand{\dens}{g~cm$^{-3}$}

\begin{document}
\title{A moderately-sized nuclear network to assist multi-D \qquad hydrodynamic simulations of supernova explosions}
%
%

\author{\firstname{Axel} \lastname{Sanz}\inst{1}\fnsep\thanks{\email{axel.sanz@estudiantat.upc.edu
    }} \and
        \firstname{Rub{\'e}n} \lastname{Cabez{\'o}n}\inst{2}\fnsep\thanks{\email{ruben.cabezon@unibas.ch
             }} \and
        \firstname{Domingo} \lastname{Garc{\'{i}a-Senz}}\inst{1,3}\fnsep\thanks{\email{domingo.garcia@upc.edu
             }}
}

\institute{Departament de F{\'{i}}sica. Universitat Polit\`ecnica de Catalunya (Spain) 
\and
           Scientific Computing Center (sciCORE). Universit{\"{a}}t Basel (Switzerland) 
\and
           Institut d'Estudis Espacials de Catalunya
          }

\abstract{A key ingredient in any numerical study of supernova explosions is the nuclear network routine that is coupled with the hydrodynamic simulation code. When these studies are performed in more than one dimension, the size of the network is severely limited by computational issues. In this work, we propose a nuclear network ({\it net87}) which is close to one hundred nuclei and could be appropriate to simulate supernova explosions in multidimensional studies.
One relevant feature is that electron and positron captures on free protons and neutrons have been incorporated to the network. Such addition allows for a better track of both, the neutronized species and the gas pressure. A second important feature is that the reactions are implicitly coupled with the temperature, which enhances the stability in the nuclear statistical equilibrium (NSE) regime.
Here we analyze the performance of {\it net87} in light of both: the computational overhead of the algorithm and the outcome in terms of the released nuclear energy and produced yields in typical Type Ia Supernova conditions.}

\maketitle
\section{Introduction}
\label{intro}
The disrupting mechanism in Type Ia Supernova (SN Ia) explosions is of thermonuclear origin and, therefore, a strong feedback between the hydrodynamics and the released nuclear energy is expected \cite{hill00}. Any hydrodynamic code aiming to simulate these explosions in more than one dimension has to incorporate good enough nuclear routines, although not too time-consuming in computing terms.

Pioneering multidimensional simulations of SN Ia incorporated around 10 nuclei with the aim of having, at best, a reasonable depiction of the released nuclear energy. For example, Benz et al. \cite{benz1989} implemented an $\alpha$-network to keep track of the nuclear evolution during the collision of two white dwarfs. Timmes et al. \cite{Timmes2000} analyzed the performance of two networks, with 7 and 14 species, concluding that they are able to account for the nuclear generation rate within $20\%$ precision. The use of these small networks has been commonplace in multi-D simulations of supernova explosions. See \cite{Reinecke99, gar99, Plewa2004}, as an example of works connected with different SN Ia explosion scenarios. At present, multidimensional simulations incorporate $\simeq 50$~nuclei, being able to reproduce the released nuclear energy within a narrow deviation, $< 5\%$~with respect a larger network \cite{gronow21}.

In this work, we describe a network of 87 species, {\it net87}, which can be used in explosive astrophysical scenarios. In particular, we applied it to assist with the explosion of a massive white dwarf (WD) made of $^{12}$C$+^{16}$O with central density $\rho_c\simeq 2\times 10^9$~\dens~in the so-called Chandrasekhar-mass explosion models. It is also adequate to track the detonation of a tiny helium shell at densities $\simeq 5\times 10^5$~\dens~located on top of a moderately massive WD, which is a necessary condition to study the SubChandrasekhar-mass route to SN Ia. 

Our proposal has two important features. First, the molar fractions of the species, $Y_i$, and the temperature, $T$, are found jointly, after implicitly solving the system of nuclear reactions coupled with the energy equation. It is well known that the implicit coupling of $\{Y_i, T\}$ leads to a much stable behavior around the NSE regime \cite{mueller86, cab04} and recent studies have identified the implicit coupling between the released nuclear energy and the nuclear reaction rates as a key point in simulating high-density combustion \citep{zin21}. The second novelty of this work is that our moderately-sized network includes the capture reactions $e^- + p \rightarrow n + \nu_e$ and $e^{+}+ n \rightarrow p + \bar\nu$. As far as we know, this is the first time that reactions of this kind are taken into account in a reduced network routine belonging to a multi-D hydrodynamic code dealing with SN Ia explosions. As we show below, the benefits of including these reactions -namely a more accurate depiction of the electronic pressure and neutronized yields- come at no relevant computational cost.

\section{Calculations and results}
\label{calculations}

\begin{figure}[h]
\noindent
\makebox[\textwidth]{
\includegraphics[width=1.1\textwidth]{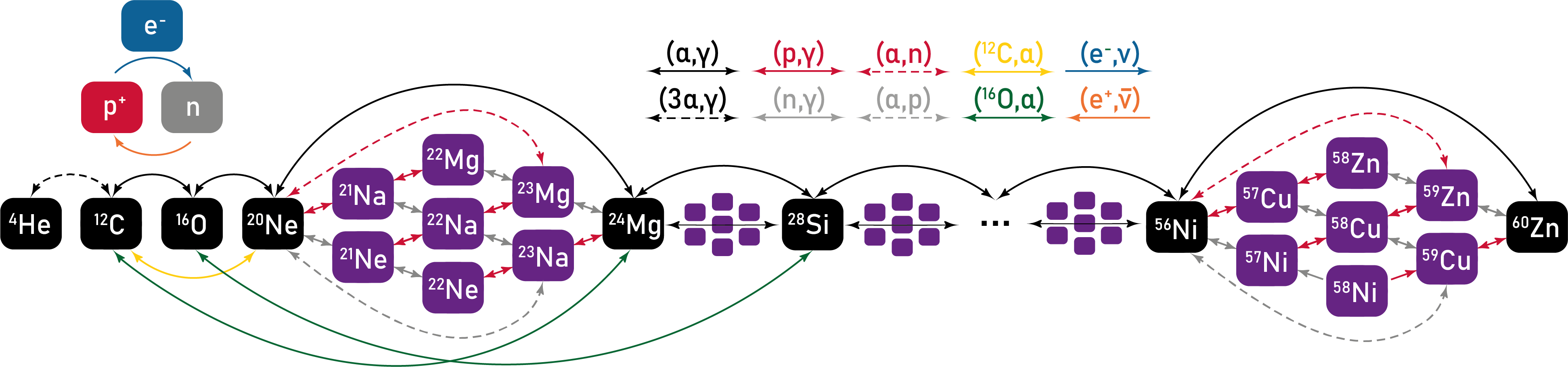}}
\caption{Schematics of the {\it net87} network, showing the $\alpha$, $p$, and $n$ reactions in the first and last cluster nodes (in purple) between $\alpha$-chain elements (in black). Three heavy-ion reactions are also included as they are the main seeds for $\alpha$ particles. Finally, the $e^-$ and $e^+$ captures are depicted on the top-left.}
\label{fig-1}       
\end{figure}

{\it net87} adds a cluster of proton and neutron reactions in between each node of a standard $\alpha-$network from $^{20}$Ne up to $^{60}$Zn (see Fig.~\ref{fig-1}). The variation of the molar fractions is integrated jointly with the temperature equation with an implicit Newton-Raphson scheme. We also included $e^-$ and $e^+$ captures on free protons and neutrons. They have well-known rates \cite{fuller1980} and may account for up to $60\%$~of bulk captures for Chandrasekhar-mass SN explosions. 

\begin{figure}
\centering
\sidecaption
\includegraphics[width=\textwidth]{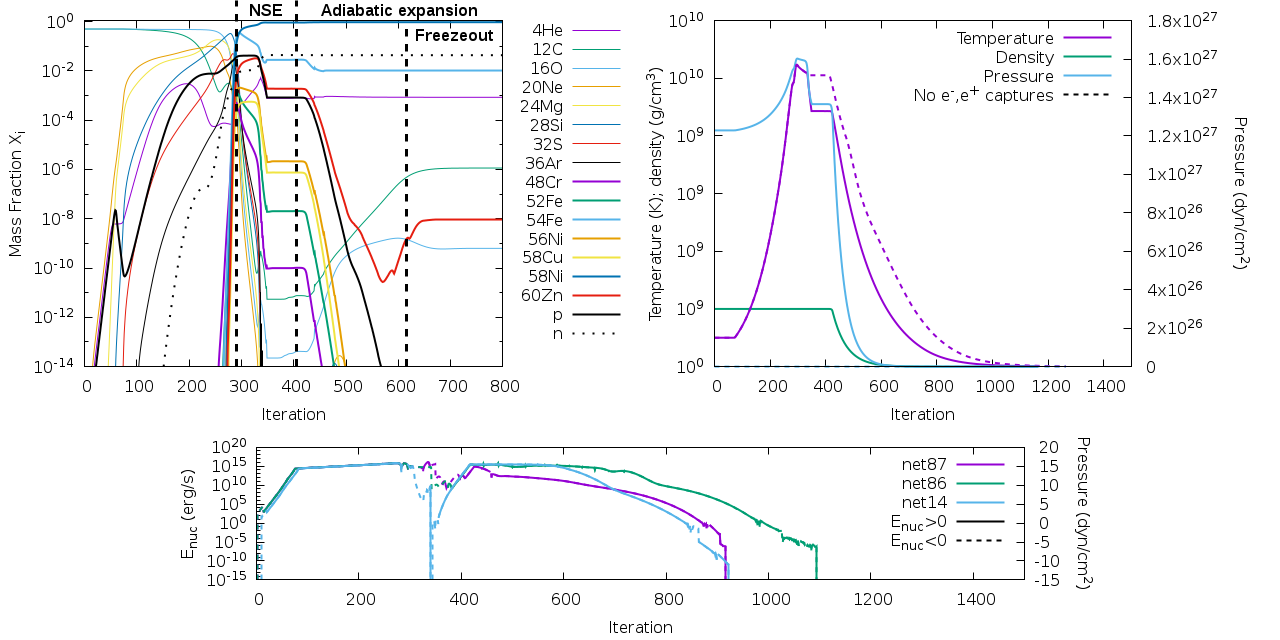}
\caption{Evolution of the most relevant quantities in the $^{12}$C$+^{16}$O test. Top-left: nuclear abundances for {\it net87}; top-right: temperature, density, and pressure for {\it net87} and {\it net86}; bottom: nuclear energy generation rate for {\it net87, net86} and {\it net14}.}
\label{fig-2}       
\end{figure}

Figure \ref{fig-2} shows the results of a $^{12}$C$+^{16}$O burning test with initial temperature $T_9=1$ and density $\rho_9=2$. An artificial decrease of density, simulating an adiabatic expansion, was switched on around iteration 400, well after NSE is established, to reach the freeze-out. Note the decrease in pressure and temperature when the endoenergetic e$^-$-captures become relevant, around iteration 300. Such reduction in pressure will have dynamical consequences in multi-D simulations of SN Ia. Additionally, the NSE plateau is well reproduced, with the abundances showing no oscillations. 

Table~\ref{tab-1} shows the computational performance of {\it net87}. The routine scales approximately as $N\log N$ with the number of species as going from {\it net14} (with 14 species) to {\it net87} rises the computing time in a factor $\simeq 15$. The overload of including $e^-$ and $e^+$ captures is negligible.

\begin{table}
\centering
\caption{ Wall-clock time, average number of Newton-Raphson iterations per global iteration, total number of global iterations, and average wall-clock time per single iteration for the $^{12}$C$+^{16}$O test and for each nuclear network \bf{{(\it net86} is {\it net87} without $e^{-}$,~$e^{+}$-captures)}}
\label{tab-1}       

\begin{tabular}{ccccc} 
\hline
        \toprule
         \multirow{2}{*}{Network} & \multirow{2}{*}{$\overline{t}$ (s)} & \multirow{2}{*}{$\overline{\textrm{NR It.}}$} & \multirow{2}{*}{It.} & $\overline{t}$/It. \\
         &&&&($\times 10^{-4}$~s)\\
         \hline
         \hline
          net14 & $0.152\pm0.009$& $4.23$ & $1792$ & $0.20$\\
          net86 & $1.92\pm0.13$  & $4.89$ & $1265$ & $3.10$\\
          net87 & $1.78\pm0.08$  & $4.87$ & $1164$ & $3.08$\\
          \hline
   \end{tabular}
\end{table}

%
%
%

\end{document}